\documentclass[twocolumn,showpacs,prl,10pt]{revtex4}
\usepackage[english]{babel}
\usepackage{amsmath,amssymb}
\usepackage{times}
\usepackage{epsfig} 

\begin{document} 
%\draft
\title{Momentum dependent relaxation rate and pseudogap in doped 
magnetic insulators}

\author{M. M. Zemlji\v c$^{1}$ and P. Prelov\v sek$^{1,2}$}
\affiliation{$^1$J.\ Stefan Institute, SI-1000 Ljubljana, Slovenia}
\affiliation{$^2$ Faculty of Mathematics and Physics, University of
Ljubljana, SI-1000 Ljubljana, Slovenia}

\date{\today}
                   
\begin{abstract}
The spectral functions and corresponding self energies are calculated
within the planar $t$-$t'$-$J$ model as relevant to hole-doped
cuprates using the exact diagonalization method at finite temperatures,
combined with the averaging over twisted boundary conditions. Results
show truncated Fermi surface at low doping and $t'<0$ in the antinodal
region while the self
energy reveals weakly ${\bf k}$- and doping dependent anomalous
relaxation rate $|\Sigma''({\bf k},\omega)|\sim a+b |\omega|$ for
$\omega<0$, consistent with recent ARPES results, and a
pseudogap-generating component of Lorentzian form. The latter is well pronounced at low
doping and strongly depends on ${\bf k}$ and $t'$.

\end{abstract}

\maketitle

It is well established that high-$T_c$ superconductivity in cuprates
develops directly from a very anomalous normal state, which does not
follow the usual Fermi-liquid (FL) scenario and thereby the concept of well
defined quasiparticle (QP) excitations. The nature of the latter is
most directly accessed by angle resolved photoemission spectroscopy
(ARPES), which probes the single-electron spectral function (SF)
$A({\bf k},\omega)$ \cite{dama}. Concentrating here on the models for
hole-doped cuprates, we furtheron discuss several ARPES normal-state
results. a) Upon doping, the insulating antiferromagnet (AFM) develops
into a metallic paramagnetic state where the Fermi surface (FS)
appears as a large one, consistent approximately with the Luttinger
FL volume in La$_{2-x}$Sr$_x$CuO$_4$ (LSCO) \cite{yosh1} or
evidently deviating from it in Ca$_{2-x}$Na$_x$CuO$_2$Cl$_2$ (Na-CCOC)
\cite{shen}. b) The FS and corresponding QP are well pronounced only
along the nodal direction at ${\bf k} \sim (\pi/2,\pi/2)$, while FS is
truncated or poorly defined towards the antinodal points ${\bf k} \sim
(\pi,0)$, where the pronounced feature is the large pseudogap first
observed in optimally-doped Bi$_2$Sr$_2$CaCu$_2$O$_{8+\delta}$
(Bi2212) \cite{mars,dama} and well developed in low-doped LSCO
\cite{ino,yosh2} and Na-CCOC \cite{shen}. c) The QP relaxation rate,
best identified along the nodal direction at optimally doped Bi2212
\cite{vall,dama}, follows the non-FL behavior $|\Sigma''({\bf
k},\omega)|\sim b(|\omega |+ \zeta T)$, as summarized within the
marginal FL concept (MFL) \cite{varm}. Recently, it
has been analyzed all along the FS and represented
by $|\Sigma''({\bf k},\omega)|\sim \tilde a_{\bf k}+ \tilde
b_{\bf k} |\omega| $, where $\tilde a_{\bf k}$ is strongly momentum
dependent and large in the antinodal part of the FS, while $\tilde
b_{\bf k}$ is nearly a constant \cite{kami}.

Above experimental facts still represent the major challenge in the
theory of correlated electrons. There are by now several numerical
results confirming that prototype models as the Hubbard model and the
$t$-$J$ model on a square lattice can account for such
phenomena. Pseudogap behavior within the Hubbard model has been
observed in the quantum Monte Carlo (QMC) studies \cite{preu}, and
more recently using dynamical cluster approximation (DCA) \cite{husc,macr}
and cellular dynamical mean-field theory (CDMFT) \cite{kyun,cive}
revealing its presence in both $A({\bf k},\omega)$ and density of states (DOS) ${\cal N}(\omega)$.
Within the $t$-$J$ model as relevant to
cuprates, the numerical study of SF using the finite temperature
Lanczos method (FTLM) established the non-FL behavior of $\Sigma''({\bf
k},\omega)$ as well as the pseudogap in DOS
\cite{jpspec,jprev}. More recent studies detected also the sensitivity
of pseudogap features to the addition of next-nearest neighbor hopping
$t'$, both in the $t$-$J$ model \cite{tohy} and the Hubbard model
\cite{cive,macr}. In spite of the accumulated evidence, there is still no
consensus on the origin of non-FL behavior and pseudogap. However,
several numerical studies \cite{preu,jpspec,jprev,macr,kyun} and also
analytical approaches \cite{prel1} suggest the short-range AFM
fluctuations as the generator.
 
In order to distinguish between different scenarios and make a
comparison with ARPES experiments, more detailed knowledge about the
momentum dependence of $\Sigma''({\bf k},\omega)$ is needed. In this
Letter we present the results for the latter, as evaluated within the
$t$-$t'$-$J$ model, improving the FTLM with a continuous ${\bf k}$
variation on small lattices. We show that the self
energy can be decomposed for $\omega\lesssim 0$ into an essentially ${\bf k}$-independent
MFL part $|\Sigma''({\bf k},\omega)|\sim a+b|\omega|$ and a
pseudogap-generating contribution, which is strongly ${\bf k}$
dependent. We also find that the pseudogap part essentially depends on
$t'$.

Our study is devoted to the extended $t$-$J$ model
\begin{equation}
H=-\sum_{i,j,s}t_{ij} \tilde{c}^\dagger_{js}\tilde{c}_{is}
+J\sum_{\langle ij\rangle}({\bf S}_i\cdot {\bf S}_j-\frac{1}{4}
n_in_j), \label{eq1} 
\end{equation}
where $\tilde{c}^\dagger_{is}= (1-n_{i,-s}) c^\dagger_{is}$ are
projected fermionic operators not allowing the double occupancy of
sites.  We include on a square lattice besides the nearest- neighbor
hopping $t_{ij}=t$ also the next-nearest-neighbor hopping
$t_{ij}=t^\prime$. We consider here $t'=-0.3 t$, as relevant
for hole doped cuprates \cite{tohy}, as well as the reference $t'=0$. In
correspondence with experiments on cuprates we also fix $J=0.3~t$
and note that $t\sim 400$~meV.

We calculate Green's function for projected fermionic
operators,
\begin{equation}
G({\bf k},\omega)= -i \int_0^{\infty}dt {\rm e}^{i(\omega+\mu )
t}\langle \{ \tilde c_{{\bf k} s}(t) , \tilde c^{\dagger}_{{\bf k} s}
\}_+ \rangle,
\label{eq2}
\end{equation}
and the corresponding SF $A({\bf k},\omega)=-\mathrm{Im}~G({\bf
k},\omega)/\pi$, where $\mu$ is the chemical potential. In the
following we present results obtained by using the exact diagonalization
approach for small tilted (Pythagorean) square lattices with
$N=n^2+m^2$ sites. Although we are primarily interested in the low
$T\to 0$ regime, there are several
advantages to perform the calculation at $T>0$, using the FTLM
\cite{jprev}.  As discussed before \cite{jpspec}, the spectra even on
a small system become quite dense and macroscopic-like at $T>T_{fs}$,
where finite-size temperature depends on the size $N$ as well as on
the model used. Consequently, the finite-size effects are substantially
reduced at $T>T_{fs}$. For our purpose it is important that in the
latter regime we are able to extract a meaningful
self-energy $\Sigma({\bf k},\omega)$ from the SF. In the following we present
results for systems of $N=18,20$ sites ranging from low to intermediate
doping, i.e. $N_h=1,2,3$ holes, where the corresponding $T_{fs}=0.1
-0.15 ~t$.  Within FTLM one evaluates separately the electron creation
and annihilation SF $A_{+,-}^{N_h}({\bf k},\omega)$ representing the
transitions $N'_h=N_h\mp 1$, respectively. In fact, we calculate
directly only $A_-^{N_h}({\bf k},\omega)$ and express $A_+^{N_h}({\bf
k},\omega)={\mathrm exp}(\beta\omega)A_-^{N_h-1}({\bf k},\omega)$. Here we
fix $\mu$ by requiring the total sum rule $(1/N)\sum_{\bf k} \int d\omega
A({\bf k},\omega) =\alpha= (1+c_h)/2$.

On a system with periodic boundary condition (BC) one is able to
consider SF only for a set of discrete ${\bf k}={\bf k_l},
l=1,N$. In order to scan the whole Brillouin zone, we
employ twisted BC by introducing the uniform vector potential $\vec
\theta$, which modifies the
hopping elements $t_{ij} \to \tilde t_{ij} = t_{ij} ~\mathrm {exp}(i
\vec \theta \cdot \vec r_{ij})$ in the Hamiltonian, Eq.~(\ref{eq1}). It is
well known that this procedure allows one to reach arbitrary momenta
${\bf k}={\bf k_l}+\vec \theta$. In the following we perform 
calculation for different phases $\vec \theta_t$, which are
chosen equidistantly in the quarter of the first Brillouin zone (FBZ). In particular,
our mesh contains $10\times 10$ and $7\times 7$ {\bf k}-points for $N=18$ and $N=20$, respectively.

Using FTLM we calculate SF \cite{jpspec,jprev} and extract
corresponding self energies $\Sigma({\bf k},\omega)$. Due to
non-trivial sum rule within the projected model, Eq.~(\ref{eq1}), the
latter are defined via
\begin{equation}
G({\bf k},\omega)= \frac{\alpha}{\omega -\zeta_{\bf k} -
\Sigma({\bf k},\omega) }. \label{eq3}
\end{equation} 
Since we require that $|\Sigma({\bf k},\omega \to \pm \infty)|\propto
1/\omega$, the 'free' term is uniquely determined as $\zeta_{\bf
k}=\int d\omega \omega A({\bf k},\omega)/\alpha$. It should be noted
that due to projection the $t$-$J$ model does not directly possess any
'free' band term therefore the dispersion $\zeta_{\bf k}$ is already
nontrivial, dependent on doping $c_h$ and model parameters
\cite{prel1}.

In order to reduce finite-size fluctuations of $\Sigma({\bf
k},\omega)$ among different ${\bf k}$ we use additional averaging
analogous to the treatment common in the cluster DMFT approaches. The
underlying idea is that $\Sigma({\bf k},\omega)$, being quite local
quantity in strongly correlated systems, varies with ${\bf
k}$ more smoothly than corresponding $A({\bf k},\omega)$. Our results
presented furtheron confirm this conjecture. Therefore, we
perform the averaging of calculated $\Sigma({\bf k},\omega)$ (as well
as $\zeta_{\bf k}$) using Gaussian weighting with radius $\delta k
\sim 0.3$. Such averaged $\Sigma({\bf
k},\omega)$ and $\zeta_{\bf k}$ are then inserted into Eq.~(\ref{eq3})
to evaluate SF.

%\vskip 0.1truecm
\begin{figure}[htb]  
\centering
\epsfig{file=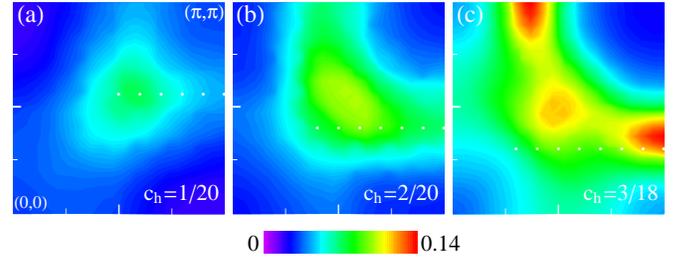,width=90mm,angle=0}
\caption{SF at the chemical potential $A({\bf k},\omega=0)$ 
in the first Brillouin zone for the $t-t'-J$ model with
$t'/t=-0.3$. Results are shown for $T/t=0.1$ and three doping concentrations. Dots
show points, where SF and $\Sigma''({\bf k},\omega)$ in Fig.~4 are
presented.}
\label{fig1}
\end{figure} 

First, we present in Fig.~1 results for the SF at the chemical
potential, $A({\bf k},\omega=0)$, calculated with $t'/t=-0.3$ at
the lowest reachable $T=0.1~t \sim T_{fs}$, as a continuous scan in the
quarter of the FBZ. As usual, such a plot is
used to locate the FS at particular doping. It is well resolved, that
at lowest doping $c_h=0.05$ (Fig.~1a) we obtain appreciable SF weight
only in the nodal direction close to ${\bf k}_K= (\pi /2,\pi /2)$ which
is clearly in agreement with experimental results for hole-doped
cuprates at low doping \cite{yosh1,shen,dama}.  Moving to the edge
of the FBZ one enters the pseudogap region, i.e. a region where the
QP peaks at the FS are strongly suppressed and the FS is in
fact not well resolvable, at least not within our numerical
limitations. While this behavior is most pronounced at the lowest $c_h$ in
Fig.~1a, it persists also for higher but still 'underdoped' $c_h=0.1$
in Fig.1b. Evidently, the FS gradually builds up with doping,
while the pseudogap remains most pronounced at the edge of the FBZ
(antinodal region). On the other hand, at 'optimum' doping $c_h \sim
0.17$ in Fig.~1c the FS becomes almost a continuous line. Moreover, at
the edge of the FBZ near ${\bf k}_X=(\pi,0)$ the SF becomes even more
pronounced than in the nodal direction, the effect indeed observed in
LSCO and Na-CCOC at intermediate doping \cite{ino,yosh1}.  A decrease in intensity
between nodal and antinodal region is, however, more reminiscent of
electron-doped Nd$_{2-x}$Ce$_x$CuO$_4$ analogue \cite{armi}. It should
be noted that the observed effective FS appears always larger than
expected from the Luttinger theorem, where $V_{FS}/V_{BZ}=
(1-c_h)/2$. A deviation is recently established also from the systematic
ARPES study of Na-CCOC \cite{shen}, whereby our results show even
larger discrepancy.

%\vskip 0.1truecm
\begin{figure}[htb]  
\centering
\epsfig{file=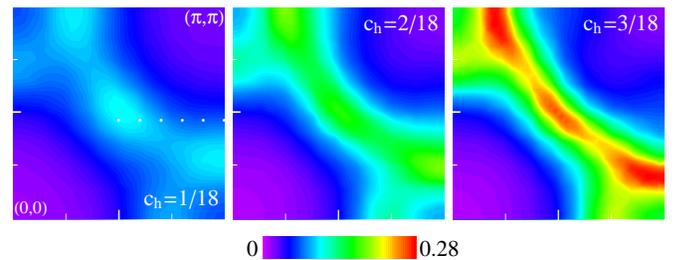,width=90mm,angle=0}
\caption{Same as Fig.~1 for the $t-J$ model.}
\label{fig2}
\end{figure}

For comparison we present in Fig.~2 also corresponding plots for the
reference $t$-$J$ model.  We detect here more continuous FS line over
the entire FBZ already in the low-doping regime. Still, the SF weight
is much weaker in the latter regime, similar to Fig.~1a. The effect of
increasing doping is clearly visible in the sharper and better defined
FS due to the increasing weight of the SF at the FS. The deviation from
the Luttinger theorem is still observable, although it is smaller relative to
Fig.~1. The difference between the SF in Figs.~1,2 can be at lowest
doping $c_h \sim 0.05$ in a naive way explained via the rigid band
picture and an effective dispersion $\epsilon({\bf k})$ of a single
hole in an AFM \cite{naza}. While within the $t$-$J$ model such a dispersion is
very anisotropic around the ${\bf k}_K$, leading to a minor difference
$\Delta \epsilon=\epsilon({\bf k}_K) - \epsilon({\bf k}_X)$,  $t'<0$
induces more isotropic dispersion around ${\bf k}_K$ and therefore
more truncated FS. On the other hand, at intermediate doping FS
already appears as a large one with the dispersion according to the
renormalized 'free' band where $t'$ directly influences its curvature.

One of the clearest manifestations of the pseudogap is the DOS ${\cal N}(\omega)=
(2/N) \sum_{\bf k} A({\bf k},\omega)$, as presented in Fig.~3, both
for $t'/t=-0.3$ and $t'=0$. We note that the results for $t$-$J$ model are
quite close to previous ones \cite{jpspec}, obtained without the BC
averaging.  This confirms that the pseudogap, more or less pronounced
at $\omega \sim 0$, is a robust feature of the model. Already from DOS
one can conclude, that pseudogap vanishes with increased doping while
$t'<0$ enhances it. E.g., for $t'<0$ DOS at $\omega \sim 0$ is reduced 
due to less coherent band and the pseudogap remains more pronounced
up to $c_h \sim 0.17$. 

\vskip 0.5truecm
\begin{figure}[htb]  
\centering
\epsfig{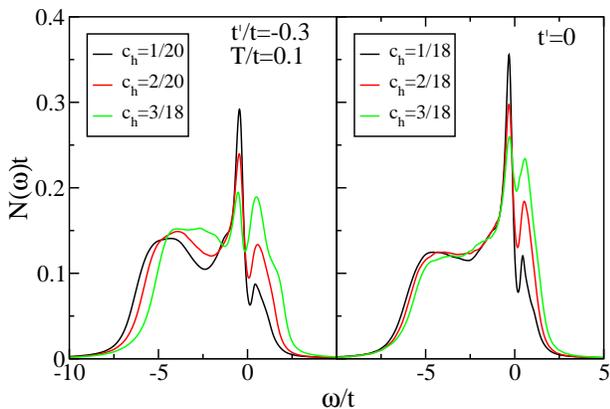}
\caption{Density of states ${\cal N}(\omega)$ at different dopings $c_h$ 
for: a) $t'/t=-0.3$ and b) $t'=0$.}
\label{fig3}
\end{figure}

However, DOS is not selective enough to reveal the origin and ${\bf k}$-dependence
of the pseudogap. Hence, we present in Fig.~4 sets of SF and
corresponding self-energies $\Sigma''({\bf k},\omega)$ taken along the
paths shown in Fig.~1 for each doping, respectively. We choose points,
being slightly below the FS in the nodal part (as this does not lead to an
essential change in $\Sigma''({\bf k},\omega)$) and matching the
pseudogapped FS at the edge of the FBZ.
%It should be again noted that the FTLM calculation of
%$A({\bf k},\omega)$ at finite $T>T_{fs}$ allows a straightforward
%extraction of $\Sigma({\bf k},\omega)$ via Eq.(\ref{eq3}), not
%requiring any additonal broadening.

Fig.~4a shows SF $A({\bf k},\omega)$ and $\Sigma''({\bf
k},\omega)$ for $t'/t=-0.3$ at lowest doping $c_h=0.05$. We notice that
along the chosen path a sharper peak in the nodal region, corresponding
approximately to a brighter spot of large weight in Fig.~1a, develops
into a pronounced two peak structure in
the antinodal region as a manifestation of the pseudogap in SF.
Besides that, SF shows well known incoherent and
nondispersive part for $\omega \ll 0$ and in this respect strong
asymmetry between $\omega<0$ and $\omega>0$ regions, well evident also in
${\cal N}(\omega)$ in Fig.~3. The latter features persist up to high
doping. In addition, from Fig.~4 it follows that the pseudogap
splitting near ${\bf k}_X$ closes on doping, whereby the QP become
better defined although still heavily damped.

\vskip -0.4truecm
\begin{figure}[htb]  
\centering
\epsfig{file=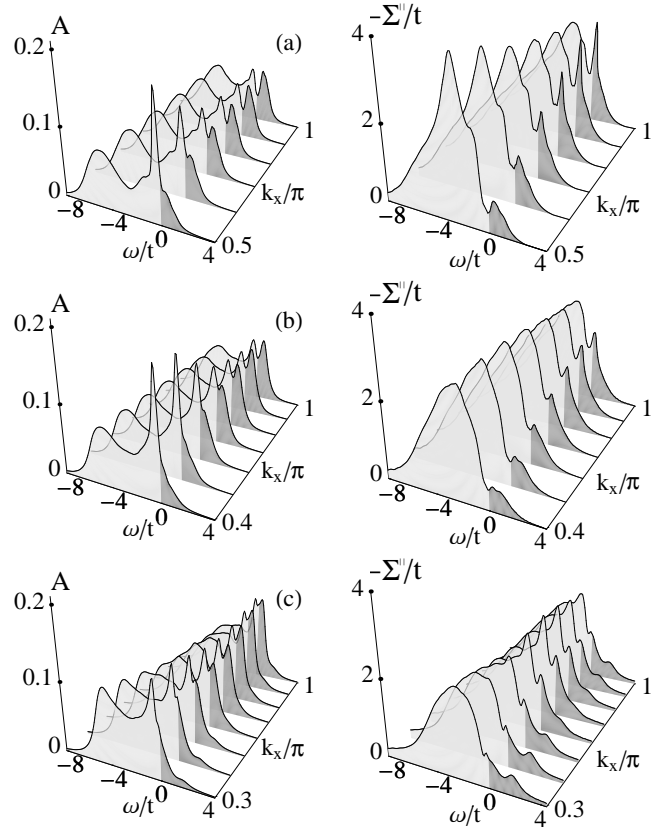,width=90mm,angle=0}
\caption{Spectral function $A({\bf k},\omega)$ (left column) and
corresponding $\Sigma''({\bf k},\omega)$ (right column) for $t'/t=-0.3$
taken in ${\bf k}$-points shown in Fig.~1 for dopings $c_h=0.05, 0.1,
0.17$, respectively.}
\label{fig4}
\end{figure}

A sensitive probe of QP damping and pseudogap behavior are self energies
$\Sigma''({\bf k},\omega)$ in Fig.~4  revealing several features: a)
at $\omega \ll 0$ large and broad damping function $|\Sigma''|\gg t$ is
consistent with the incoherent part of SF, b) much smaller damping for $\omega
\gg 0$, c) linear behavior $\Sigma''({\bf k},\omega \lesssim 0)$ as
described within the MFL scenario, d) a pronounced peak of a Lorentzian
form is building-up at $\omega\sim 0$ as we move to the antinodal part
of FS, representing a clear signature of the pseudogap. Namely, a sharp
peak $|\Sigma''|\propto \pi \Delta^2 \delta(\omega)$ would open a real
gap in the SF at the FS, while broader one opens the pseudogap, with the
effective peak splitting $E_{PG} \sim 2\Delta$.

%\vskip 0.7truecm  
\begin{figure}[htb]   
\centering
\epsfig{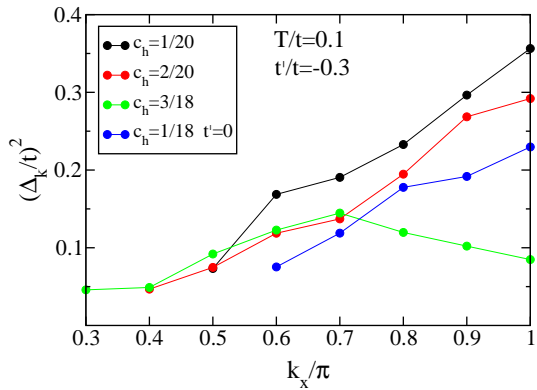}
\caption{Momentum dependence of the pseudogap weight $\Delta_{\bf k}^2$
obtained for $t'/t=-0.3$ and $t'=0$ for different
$c_h$ along the paths presented in Figs.~1,2.}
\label{fig5}
\end{figure}

It follows from Fig.~4 that the pseudogap contribution in $\Sigma''({\bf k},\omega)$ for $t'/t=-0.3$
is largest in the antinodal region and for lowest doping $c_h=0.05,
0.1$, where it is also very sharp. Elsewhere it loses weight and
becomes indistinguishable from the background. In order to quantify $\Sigma''({\bf
k},\omega)$ close to FS and to study its momentum dependence we assume
for $\omega\lesssim 0$ the form
\begin{equation}
 \Sigma({\bf k},\omega) \sim \Sigma_{MFL}({\bf k},\omega)
+\Delta^2_{\bf k}/(\omega-\omega^*_{\bf k}+i \Gamma_{\bf k})
.\label{eq4}
\end{equation}  
whereby the MFL part is taken as $-\Sigma_{MFL}''({\bf k},\omega \sim 0
) \sim a_{\bf k} + b_{\bf k}|\omega|$. We use Eq.~(\ref{eq4}) to fit
the calculated $\Sigma''({\bf k},\omega)$ within the range
$-t<\omega<0.2 t$. The latter is chosen in this way since at $\omega<0$ the
linear part extends at least down to $\omega \sim -t$ (at $c_h=0.17$
even to $\omega \sim -2t$) while the position of the peak $\omega^*_{\bf k}
<0.2 t$ and the linear part at $\omega>0$ is also very
restricted. Moreover, the fit and the pseudogap interpretation is
meaningful only as far as $\Gamma_{\bf k} <\Delta_{\bf k}$.

The fitting analysis using Eq.~(\ref{eq4}) yields results presented in Fig.~4.
Again, the
variation along the paths shown in Figs.~1,2 is considered, whereby only points
with $\Gamma_{\bf k}<\Delta_{\bf k}$ are presented.  We observe for
$t'/t=-0.3$, as expected from Fig.~3, a clear decrease of
$\Delta_{\bf k}$ towards the nodal direction for $c_h=0.05, 0.1$,
whereas for 'optimum' $c_h=0.17$ the weight is reduced in particular
in the antinodal region with a maximum at an intermediate ${\bf k}$
along the path. The $t$-$J$ model reveals smaller $\Delta_{\bf k}$, in
particular it is hardly resolved for larger $c_h =0.11, 0.17$, hence
we omit these data in Fig.~4.

As expected, $b_{\bf k}$ shows much less pronounced momentum and
even doping dependence. The linear regime is particularly broad
for the 'optimum' doping $c_h=0.17$, where we find $b_{\bf
k}=1.45\pm 0.2, 1.2 \pm 0.2$ for $t'/t=-0.3$ and $t'=0$,
respectively. In order to make a quantitative comparison with
experimentally determined $\tilde b \sim 0.8$ \cite{kami} one can compare only the
effective damping $Z_{\bf k} |\Sigma''({\bf k},\omega)|$ whereby $Z_{\bf
 k}$ is the QP weight well defined only in the region without pseudogap. In
ARPES, $Z_{\bf k}$ is less clearly defined and is taken $Z_{\bf k} \sim
0.5$ \cite{kami}, while we extract $Z_{\bf k} \sim 0.3\pm 0.1$. This brings
model $b_{\bf k}$ and experimental $\tilde b$ close together.

Similar results we get for lower $c_h \leq 0.1$, i.e.  $b_{\bf
k}=1.8\pm 0.2, 1.5 \pm 0.2$ for $t'/t=-0.3$ and $t'=0$, respectively.
However, the range of linearity shrinks for low doping compatible with
less justified MFL form for $\omega
\sim 0$. Furtheron, the MFL scenario predicts $a \sim \pi b T$ \cite{varm}. As our
results are calculated at $T=0.1~t$, we expect the MFL contribution $a
\sim 0.5~t$. For the intermediate doping $c_h=0.17$ and $t'=-0.3~t$ 
we indeed find $a_{\bf k}$ close to this value along the nodal
direction, with the increasing tendency towards the edge of the FBZ
where $a_{\bf k}\sim t$.  This is consistent with the analysis of ARPES in
Bi2212 where in the antinodal region $\tilde a_{\bf k} \sim 200$~meV has been
observed \cite{kami}.  However, within our approach a more detailed
analysis of $a_{\bf k}$ is difficult since it is hard to
separate it from the pseudogap contribution, in particular at lower
doping, where the pseudogap part dominates. Furthermore, our results indicate that $a_{\bf k}$ and
the pseudogap contribution could become the same feature when $\Gamma > \Delta$. Similar
question can arise also in the interpretation of Bi2212 data where the
pseudogap appears to be present as well \cite{kami}.

In conclusion, we have shown that within the $t$-$t'$-$J$ model as
relevant for hole-doped cuprates the self energy $\Sigma({\bf
k},\omega)$ can be decomposed into two parts. $\Sigma_{MFL}({\bf
  k},\omega)$ is almost ${\bf k}$-independent
as well as weakly changing with doping and $t'$, but still
anomalous and following the MFL scenario. Another contribution generating
the pseudogap is strongly ${\bf k}$-dependent most pronounced in the
antinodal part of the FBZ and at low doping. It is much more
expressed for $t'<0$ case, which is more relevant for
hole-doped cuprates. It seems that at intermediate doping the
pseudogap contribution transforms
into a constant part of QP damping $a_{\bf k}$ which therefore becomes ${\bf k}$-dependent. 
 
It should be noted that the pseudogap scale discussed here is the
large (high-energy) one \cite{dama} as e.g. very evident in LSCO \cite{ino,yosh1}
and Na-CCOC \cite{shen} at low doping. The latter is responsible
for effective truncation of the FS in this regime, as observed
in our results in Figs.~1,2. We also emphasize that our analysis
as well as experimental does not give a clear answer whether such a pseudogap
indeed vanishes in the nodal direction, or the weak coherent QP peak
appears inside the pseudogap. As far as the origin of these
phenomena is concerned, ${\bf k}$ and $t'$ dependence of the pseudogap
as well as underlying $\Sigma_{MFL}(\omega)$ are in favor of the
interpretation in terms of coupling to short-range AFM
spin fluctuations \cite{prel1,kyun,macr}.

We acknowledge useful discussions with T. Tohyama.
This work was supported by
the Slovenian Research Agency under grant Pl-0044.


\begin{thebibliography}{99}                                          

\bibitem{dama} for a review see A.\ Damascelli, Z.\ Hussain, and
Z.-X.\ Shen, Rev.\ Mod.\ Phys. \textbf{75}, 473 (2003).
\bibitem{yosh1} T.\ Yoshida {\it et al.}, cond-mat/0510608.
\bibitem{shen} K.\ M.\ Shen {\it et al.}, Science \textbf{307},
910 (2005).
\bibitem{mars} D.\ S.\ Marshall {\it et al.}, Phys.\ Rev.\
Lett. \textbf{76}, 4841 (1996).
\bibitem{ino} A.\ Ino {\it et al.}, Phys.\ Rev.\
Lett. \textbf{81}, 2124 (1998);  A.\ Ino {\it et al.}, Phys.\ Rev. B \textbf{62}, 4137
(2000).
\bibitem{yosh2} T.\ Yoshida {\it et al.}, Phys.\ Rev.\
Lett. \textbf{91}, 027001 (2003).
\bibitem{vall} T.\ Valla {\it et al.}, Science \textbf{285}, 2110
(1999).
\bibitem{varm} C.\ M.\ Varma, P.\ B.\ Littlewood, S. Schmitt-Rink, E.\ 
Abrahams, and A.\ E.\ Ruckenstein, Phys.\ Rev.\ Lett.\ \textbf{63},
1996 (1989).
\bibitem{kami} A.\ Kaminski {\it et al.}, Phys.\ Rev.\ B \textbf{71},
014517 (2005).
\bibitem{preu} R.\ Preuss, W.\ Hanke, C.\ Gr\" ober, and
H.\ G.\ Evertz, Phys.\ Rev.\  Lett. \textbf{79}, 1122 (1997).
\bibitem{macr} A.\ Macridin, M.\ Jarrell, Th.\ Maier, and P.\ R.\ C.\ Kent, cond-mat/0509166.
\bibitem{husc} C.\ Huscroft, M.\ Jarrell, Th.\ Maier, S.\ Moukouri and
A.\ N.\ Tahvildarzadeh, Phys.\ Rev.\ Lett. \textbf{86}, 139 (2001).
\bibitem{kyun} B.\ Kyung, S.\ S.\ Kancharla, D.\ Senechal, A.\ - M.\
S.\ Tremblay, M.\ Civelli, and G.\ Kotliar, Phys.\ Rev.\ B
\textbf{73}, 165114 (2006).
\bibitem{cive} M.\ Civelli, M.\ Capone, S.\ S.\ Kancharla, O.\
Parcollet, and G.\ Kotliar, Phys.\ Rev.\ Lett. \textbf{95}, 106402
(2005).
\bibitem{jprev} for a review see J.\ Jakli\v c and P.\ Prelov\v sek,
Adv.\ Phys.\  \textbf{49}, 1 (2000).
\bibitem{jpspec} J.\ Jakli\v c and P.\  Prelov\v sek, Phys.\ Rev.\ B 
\textbf{55}, R7307 (1997);  P.\ Prelov\v sek, J.\ Jakli\v c, and K.\ Bedell,
Phys.\ Rev.\ B \textbf{60}, 40 (1999).
\bibitem{tohy} T.\ Tohyama, Phys.\ Rev.\ B \textbf{70}, 174517 (2004). 
\bibitem{prel1} P.\ Prelov\v sek and A.\ Ram\v sak,  Phys.\ Rev.\ B 
\textbf{63}, 180506(R) (2001).
\bibitem{armi} N.\ P.\ Armitage {\it et al.}, Phys.\ Rev.\
Lett. \textbf{88}, 257001 (2002).
\bibitem{naza} A.\ Nazarenko, K.\ J.\ E.\ Vos, S.\ Haas, E.\ Dagotto,
  and R.\ J.\ Gooding,  Phys.\ Rev.\ B 
\textbf{51}, 8676 (1995).

\end{thebibliography}
\end{document}